# Realization of A Non-Markov Chain in A Single 2D Crystal RRAM

Rongjie Zhang[1,#], Wenjun Chen[1,#], Changjiu Teng[1], Wugang Liao[2], Bilu Liu[1*] and Hui-Ming Cheng[1,3*]

[1] *Shenzhen Geim Graphene Center, Tsinghua-Berkeley Shenzhen Institute and Tsinghua Shenzhen International Graduate School, Tsinghua University, Shenzhen, 518055, P. R. China*

[2] *Institute of Microscale Optoelectronics, Shenzhen University, Shenzhen 518060, P. R. China*

[3] *Shenyang National Laboratory for Materials Science, Institute of Metal Research, Chinese Academy of Sciences, Shenyang 110016, P. R. China*

[#]These authors contributed equally to this work.

Email: bilu.liu@sz.tsinghua.edu.cn (BL), hmcheng@sz.tsinghua.edu.cn (HMC)

**Abstract**

The non-Markov processes widely exist in thermodymanic processes, while it usually requires packing of many transistors and memories with great system complexity in traditional device architecture to minic such functions. Two-dimensional (2D) material-based resistive random access memory (RRAM) devices show potential for next-generation computing systems with much-reduced complexity. Here, we achieve the non-Markov chain in an individual RRAM device based on 2D mica with a vertical metal/mica/metal structure. We find that the internal potassium ions ($K^+$) in 2D mica gradually move along the direction of the applied electric field, making the initially insulating mica conductive. The accumulation of $K^+$ is tuned by electrical field, and the 2D-mica RRAM possesses both unipolar and bipolar memory windows, high on/off ratio, decent stability and repeatability.

Importantly, the non-Markov chain algorithm is established for the first time in a single RRAM, in which the movement of $K^+$ is dependent on the stimulated voltage as well as their past states. This work not only uncovers the inner ionic conductivity of 2D mica, but also opens the door for such novel RRAM devices with numerous functions and applications.

**Keywords:** 2D materials, mica, ion transport, RRAM, non-Markov chain.

**Introduction**

Information technology plays increasingly important roles in human society while conventional von-Neumann architecture is approaching its limit when processing the exponentially growing amount of data.[1,2] Combining memory/sensor cells and algorithms into cutting edge hardware and devices[3] is important in next-generation computing.[4,5] To this end, resistive random access memory (RRAM) or memristors[6,7] have been put forward to compliment the traditional non-volatile memory devices owing to their high operation speed, high integration density, as well as applications in sophisticated condition.[8–11] Recently, two-dimensional (2D) materials have emerged as a versatile platform for RRAM applications.[12–17] Compared with bulk materials, the RRAM devices based on 2D materials possess outstanding performance. For example, 2D material-based RRAM devices have demonstrated ultra-fast switching of < 10 ns,[18] the thinnest nonvolatile resistive switching (RS),[19] sub-pA switching current,[20] ultralow operation voltage of 100 mV,[21] high-frequency operation at GHz level,[22] and excellent stability at high temperature of up to 340 °C.[23]

The high integration of RRAMs is advantageous to accomplish specific functional algorithms in a machine learning system. Although deep learning algorithms can be realized by software, numerous transistors are usually needed,[24] which may degrade the operation stability and robustness of the system. Several algorithms, such as sparse coding, convolution neural network and reinforcement learning, are accomplished by using RRAM arrays.[25–30] The application of RRAM arrays requires software and hardware assistance. To overcome these limitations, the Markov chain, which is usually realized by software, was achieved in a single RRAM based on 2D SnSe.[31] The Markov chain simplifies and reduces the calculation in the system, which shows great potential in machine learning and automatic speech recognition. However, various physical phenomena are related to the non-Markov[32] property, where the probability of the next state is related to both current and past states. For instance, the Markov chain is not applicable in the systems which are related with their historical states, such as time-dependence human neural memory brain, environmental-related quantum entangled state,[33] thermodynamic processes, and cumulative effect, etc. However, the achievement of a hardware non-Markov chain is elusive as it refers to complicated process, which hindrads the application of non-Markov process in historical related physical processes. In addition, the vacancy-based or external ion based RS mechanism may cause the variation of RRAM performance, such as random telegraph noise[34] and set voltage variability,[35] which is challenging to be addressed. Therefore, it is urgent to explore new RS mechanism to achieve non-Markov process with high performance.

Here, we report the realization of a non-Markov chain in a single 2D crystal RRAM with a novel internal ionic conduction mechanism. We find that the migration of the internal $K^+$ inside 2D mica leads to the formation of conductive filament (CF) paths, which is comfirmed by electrical measurements and the depth anylysis of $K^+$. The 2D mica RRAM exhibits both unipolar and bipolar conductive behaviors with a high on/off current ratio of $10^3$, high stability and repeatability, verified by conductive atomic force microscopy (CAFM) measurements. The mica-based RRAM is capable to remember the polarity of input voltages, resulting in the different response behaviors of $K^+$ in mica under positive and negative electrical fields. Accordingly, it is the first time to realize the non-Markov process in an individual RRAM device. The internal ion transport demonstrated in this paper provides additional dimensions in memory devices, opening a new avenue in the design and fabrication of intelligent devices.

**Results and discussion**

Figure 1a shows the structure and measurement schematics of the RRAM based on 2D mica, in which the AFM tip is an electrode to apply voltage and the bottom Au electrode is another one. The application of electric field leads to the accumulation of $K^+$ along the direction of the electrical field, which forms the CF in the vertical direction of mica (Figure 1b). Here, trilayer mica with a thickness of 3 nm is selected to reveal the RS behaviors. Figure 1c shows the optical image of the mica flake, and its thickness is determined by AFM (inset of Figure 1d). The current-voltage (*I-V*) curves (Figure 1e) displays that the mica-based device is initially insulating then conductive under the application of a cyclic triangular wave voltage with the maximum value of 2 V. The typical unipolar RRAM window is observed after 30 swept cycles, representing the formation of CF in the device. The mica is initially insulating (with a resistance >10 GΩ), and gradually switches to a low resistance state (< 0.1 GΩ) at a positive voltage of 1.5 V. The formation of CF results in stable RRAM features in the following sweeps. When the amplitude of the ramped voltage is increased to 4 V, the mica exhibits stable bipolar RRAM windows after 10 swept cycles (Figure 1f), indicating the easier formation of CF in mica under a higher electrical field. On the contrary to the traditional RRAM with either unipolar or bipolar windows, the mica-based device uniquely combines unipolar and bipolar performance, the transformation between which can be simply tuned by controlling intensity of the applied electric field.

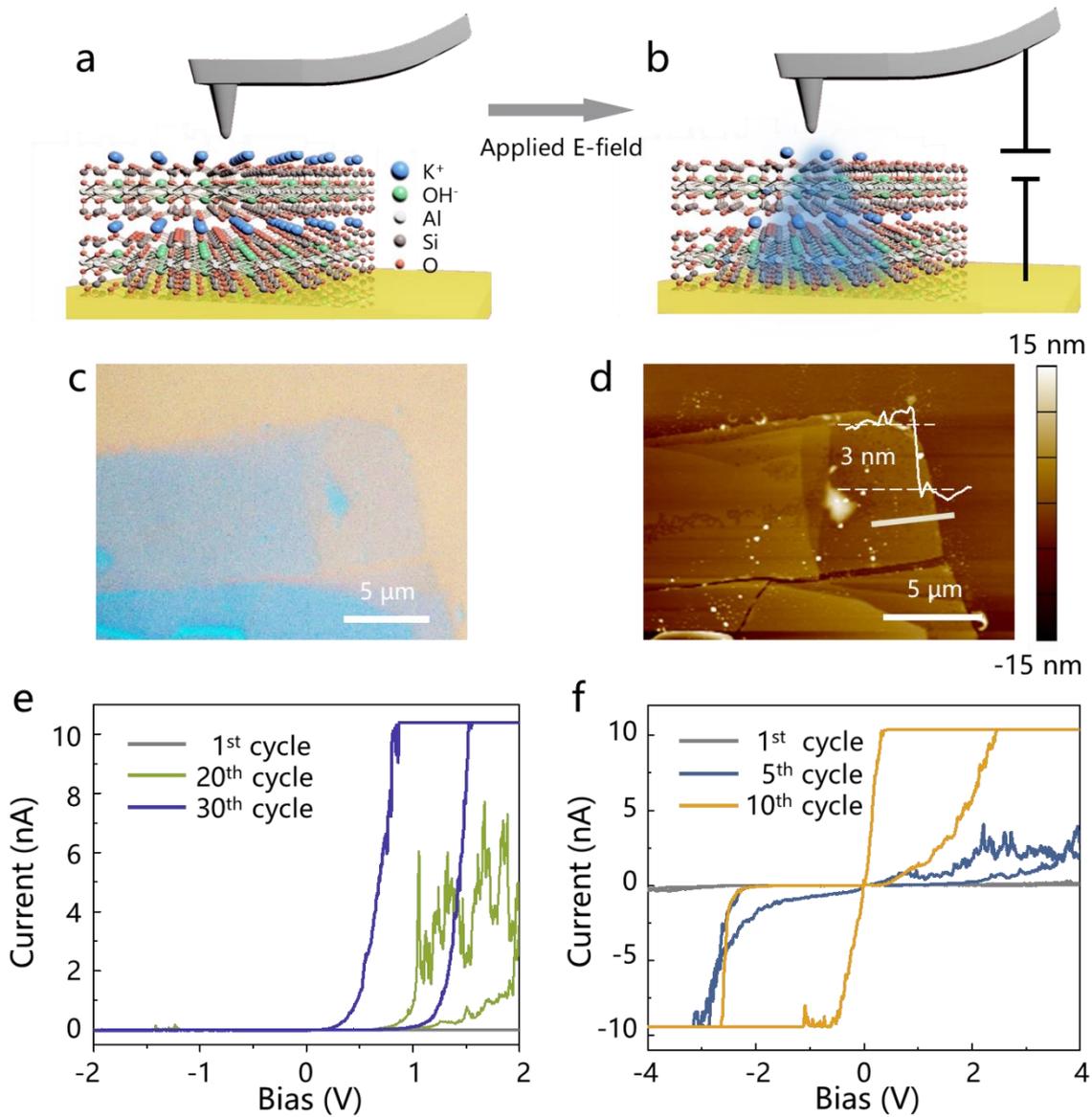

**Figure 1.** Device structure and electrical characteristics of 2D-mica-based RRAM. Schematic of the distribution of $K^+$ in the device (a) before and (b) after the application of a positive electric field. (c) Optical microscopy image of 2D mica on Au/SiO$_2$/Si substrate. (d) AFM image of the thin-layer mica in (c). Inset is the height analysis of the mica, showing its thickness of ~3 nm. (e) The 1$^{st}$, 10$^{th}$, and 30$^{th}$ RS characteristics of the mica-based device at a cyclic triangular wave voltage with the amplitude of 2V. (f) The 1$^{st}$, 5$^{th}$, and 10$^{th}$ RS characteristics of the mica-based device at a cyclic triangular wave with the amplitude of 4V.

The accumulation and migration of $K^+$ driven by an electric field are proposed to reveal the conductive behaviors of mica. To verify the electrodialysis ability of $K^+$ in mica, a device based on aqueous potassium iodide (KI)/mica/Au with free $K^+$ was fabricated (Figure 2a). In detail, a layer of photoresist (PR) was spin-coated and dried to cover the whole substrate. Then a hole was made in the PR layer by laser ablation that serves as a window to allow the external $K^+$ offered by KI/DI water solution to move through the mica (see details in methods). The *I-V* curves of the device with one probe attached to the bottom Au electrode (ground) and another immersed into KI solution droplet, were measured in air. As shown in Figure 2b, the KI/mica/Au device is initially at a high-resistance state (>1 GΩ), and transformed into a low-resistance state (<200 kΩ) after the application of positive voltage with a maximum value of 20 V. The reason is that the $K^+$ in KI solution would gather to the top of mica and then gradually migrate through the $Al_2(AlSi_3O_{10})(OH)_2^-$ layer in mica, which causes the conduction of mica. To further confirm the migration mechanism of $K^+$, a pulse voltage (±2.5 V) was used to stimulate the KI/mica/Au device after the formation of CF. A clear current vibration at the voltage of 1 V is exhibited in 150 pulse cycles (Figure 2c). The conductivity of the KI/mica/Au device is decreased under the negative pulse voltage, since the $K^+$ is forced to move back to the top layer of mica. On the contrary, the device becomes more conductive under the positive voltage because of the continuous accumulation of $K^+$. The results reveal that the external $K^+$ can transport and migrate in the out-of-plane direction of mica under the electrical field.

In addition to the external $K^+$, the accumulation and rearrangement of the internal ones in mica also contribute to the RS behaviors. To confirm the migration of internal $K^+$ in mica, a field effect transistor (FET) with mica as the dielectric, $MoS_2$ as the channel, and graphite as the electrodes, was fabricated. The migration of $K^+$ in mica plays a role in gating effect due to their positive charge (Figure S1). As a result, the $MoS_2$ FET shows synaptic responses at the gate pulse voltage, which indicates the accumulation and rearrangement of the internal $K^+$ in 2D mica (Figure S2). Moreover, another device based on a vertical graphite/mica/graphite structure (Figure S3a) was fabricated to rule out the possibility that the formation of CF from the electrodes. Figure S3b displays a similar RRAM window of the graphite/mica/graphite device. It is known that the carbon atoms in graphite are immovable under the electrical field because of the strong covalent bonds. From all the above results, we conclude that the vertical conduction of mica is ascribed to the movement of its internal ions.

Next, we analysis the conduction mechanism of the device by examing the concentrations of $K^+$ at different stages. We use time-of-flight secondary-ion mass spectrometry (TOF-SIMS), a common technique for depth profiling of materials[36], to explore the migration of $K^+$ in mica. Here, we study the graphite/mica/graphite structure (Figure 2d) with TOF-SIMS. Figure 2e shows the profiles of C, Si, K, and Al elements as a function of the depth in the device. The two separated regions of C signal are corresponding to the top and bottom graphite, respectively, between which the signals of Si, K, and Al belong to the mica layer sandwiched by two graphite electrodes. Before the application of voltage, the intensity of K keeps nearly constant, suggesting the homogeneous distribution of $K^+$ in mica. After the application of a positive vertical voltage (Figure 2f), the increasing gradient of K profile reflects the movement of $K^+$ from the top of mica toward the bottom, as schemed in Figure 2a. Therefore, the TOF-SIMS analyses reveal that the conduction of mica in out-of-plane direction relies on the migration of $K^+$ under electrical field.

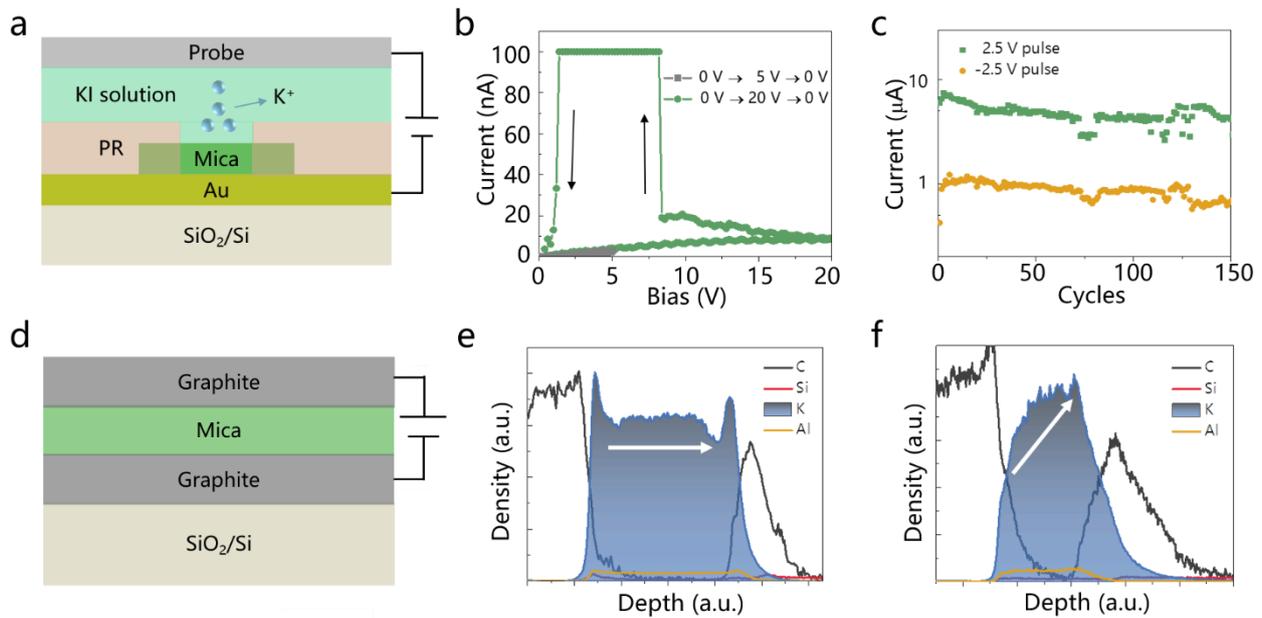

**Figure 2.** $K^+$ transport along the vertical direction in 2D mica. (a) Schematic of the probe/KI solution/mica/Au device for the measurements of $K^+$ transport. RS characteristics (b) of the device in (a). (c) The current in a read voltage of 1 V of the device in (a) after the application of voltage pulses ($\pm 2.5$ V). (d) Schematic of the graphite/mica/graphite device. The TOF-SIMS depth profiling of C, Si, K and Al elements in graphite/mica/graphite (e) without and (f) after the application of a positive voltage.

The unique RRAM behaviors in 2D-mica-based devices can be understood by the migration of $K^+$ under electrical field. Figure 3a exhibits the counterclockwise unipolar RRAM window of the device aroused by a cyclic triangular wave voltage with an amplitude of 2 V. Interestingly, the unipolar window is transformed to a clockwise bipolar one when the amplitude of the periodic voltage is increased to 5 V (Figure 3b). The formation of two distinguished RRAM windows is mainly attributed to the scanning of voltage at different ranges. Figure 3c reveals the dynamics of $K^+$ after the device was stimulated by a triangular wave voltage with the amplitude of 2 V. The $K^+$ from the top layer of mica go through the $Al_2(AlSi_3O_{10})(OH)_2^-$ layer under the 0 V to 2 V voltage sweep. The local accumulation of $K^+$ enables the conduction of mica, which causes the increasing current in the vertical direction of mica. The subsequent positive electric field (2 V to 0 V sweep) continuously motivates the gathering of the positive ions. Thus, the mica is more conductive at this stage than the former, resulting in the RRAM window. Noteworthily, the accumulated $K^+$ are regressed to the initial position immediately once the electrical field is changed to negative, which turns the mica to the insulation state. Consequently, the device shows unipolar RRAM windows under the ramped voltage sweep from -2 V to 2 V. As a contrast, in the case of higher electrical field (cyclic triangular wave voltage sweep with the amplitude of 5V), the mica-based RRAM shows bipolar behavior, the mechanism of which is schemed in Figure 3d. The higher negative voltage turns over the arrangement of $K^+$ compared with the positive electrical filed. The accumulation of $K^+$ connects the two electrodes as well, contributing to the conduction of mica. As a result, the mica-based RRAM exhibits low resistance state in 0 V to 5 V sweep. The accumulation of $K^+$ would gradually disperse under the positive electrical field, which provides enough energy to reinstate mica to its original insulation state. That makes the device exhibit a clockwise RRAM window in the positive voltage. From 0 V to -5 V sweep, the $K^+$ slightly re-accumulate under higher input electrical field. Further application of negative voltage produces higher current in mica than before with the increasing accumulation of $K^+$, leading to the RRAM window in negative voltage. In brief, the $K^+$ in mica migrate through the $Al_2(AlSi_3O_{10})(OH)_2^-$ layer under electrical field, forming the CF in the mica-based RRAM. The different accumulation directions of $K^+$ under positive or negative electrical field contribute to unipolar or bipolar RRAM windows in the mica-based RRAM.

Besides RRAM windows, repeatability and on/off ratio are two important criteria to evaluate the performance of RRAM. As indicated in Figure S4a and b, both the unipolar and bipolar *I-V* curves keep stable during ten cycles of triangular wave voltage, illustrating the repeatability of the RRAM behaviors. In addition, a high on/off ratio of $10^3$ is obtained, which is sufficient to realize the functions of a RRAM (> 10).[37] Apart from 3-nm-thick mica, similar phenomena (Figure S5a and b) were also observed for the devices based on mica with the thickness of 4.1, 6.3 and 7.9 nm. We find that the thicker the mica is, the higher the sweep voltage is needed to form the CF. To avoid electrical performance variations among samples, more than three mica flakes in each thickness were measured and they exhiti reproducible behaviors. Figure 3e exhibits the storage time of the mica-based RRAM after set and reset process by pulse voltage. The obvious resistance difference is observed after measuring the device for 800 s, with a current on/off ratio around $10^2$, indicating that the migration of $K^+$ can retain the accumulated state persistently. Taking advantage of the high on/off ratio, three memory states are demonstrated by use of the pulse voltage on the mica-based RRAM. Figure 3f shows the current of the device at a read voltage of 1 V after each pulse. The device shows distinct current states in 200 cycles. The low resistance (<0.1 GΩ), high resistance (0.1-1 GΩ) and insulating state (>1 GΩ) were achieved through the application of +10 V pulse, +5 V pulse and -10 V pulse, respectively. Furthermore, a device arry with 8 × 8 positions in the 2D mica flake are selected (Figure S6a). The current of all points in the array are below 10 pA under the read voltage of 1 V, indicating the high uniformity in resistance at the initial state (Firuge S6b). In comparison, the current increases to the compliance value (10 nA) after the application of a 10 V voltage to set some positions of the array. According to the current differences, the patterns of "T", "B", "S", and "I" are achieved in the array (Figure 3g). These results show that the mica-based RRAM displays high stability, repeatability, reproducibility and multi-state storage.

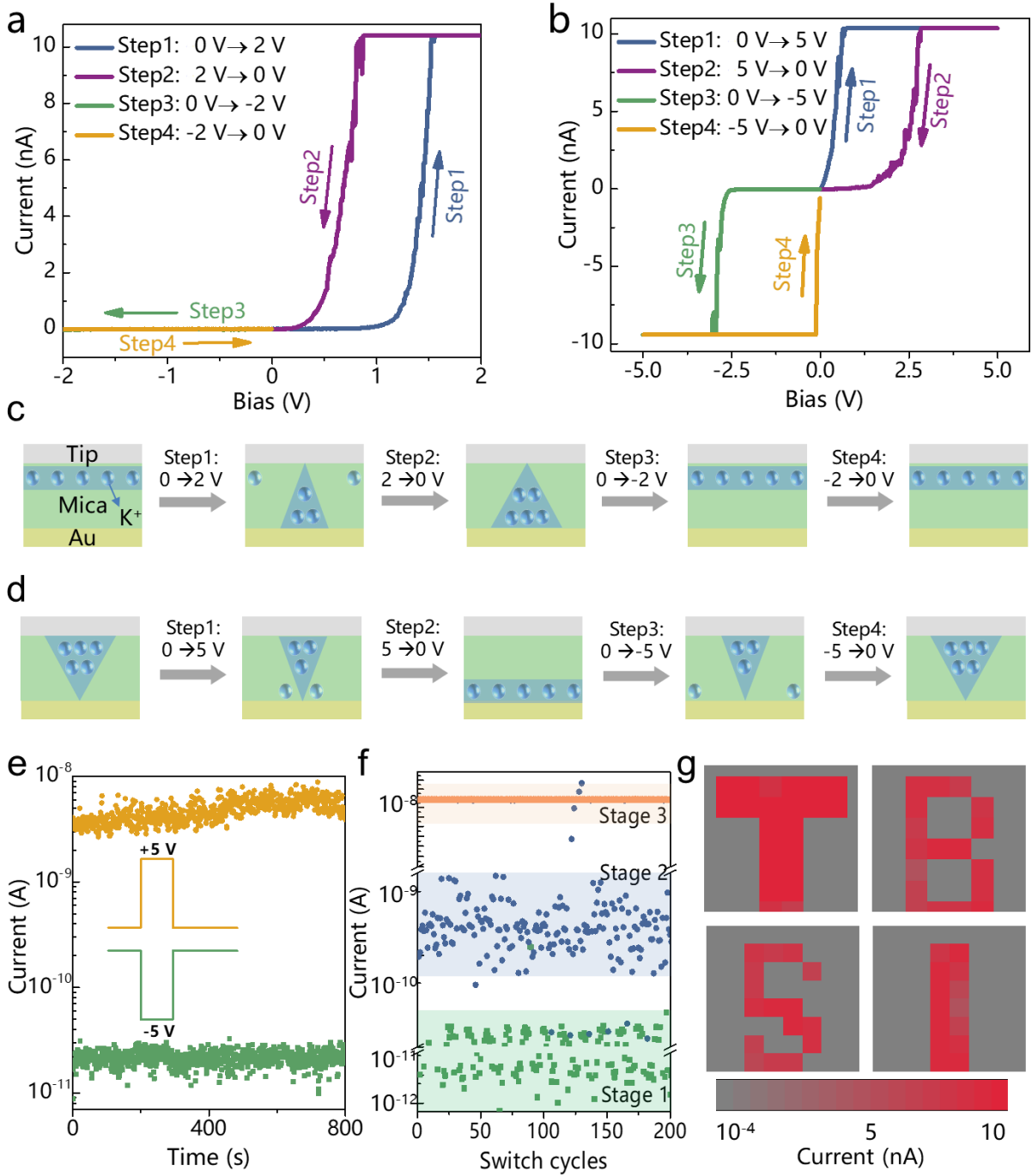

**Figure 3.** Electrical performance of the 2D-mica-based RRAM. (a) Unipolar RRAM behavior of the device at the stimulation of periodic triangular wave voltage with the amplitude of -2 to 2 V. (b) Bipolar RRAM behavior of the device at the stimulation of periodic triangular wave voltage with the amplitude of -5 to 5 V. Schematic of the accumulation of $K^+$ in mica under the periodic triangular wave voltage sweep with the amplitude of (c) 2 V and (d) 5 V. (e) The current evolution of the set and reset states in the device after 800s. (f) The current distribution of more than 200 switching cycles of three memory states. The current in (e) and (f) were read at a voltage of 1 V. (g) Pattern of a device array with 8 × 8 positions in 2D mica measured under 1 V, in which the composed areas of "T", "B", "S", and "I" patterns are pre-set to be on with a pulse voltage of 10 V.

Interestingly, the state maintenance of $K^+$ in mica enables the related device to exhibit the property of a non-Markov chain, which can be considered for the relevant applications based on the non-Markov process. The typical non-Markov process can be described as:[31]

$P(x_{n+1} = j | x_n = i_n, x_{n-1} = i_{n-1}, x_{n-2} = i_{n-2}, ..., x_0 = i_0,) \neq P(x_{n+1} = j | x_n = i_n)$, where P is the probability of the states, $x_{n-1}, x_{n-2}, ..., x_0$ are the past states, $x_n$ is the current state, and $x_{n+1}$ represents the next state of the non-Markov chain. Figure 4a shows the transfer matrix of a typical non-Markov chain, in which the transfer probability in each state is a function of both current and past states. In our design, three non-Markov states are defined, the insulating (State I), high resistance (HR) (State II) and low resistance (LR) (State III), according to the resistance of the mica-based RRAM. The cartoons in Figure 4b exhibit parts of the transfer paths of the three states in the device. The positive and negative voltages are applied to imitate different transformations of states in the non-Markov chain. Two steps of input voltages are applied to inspire the device, in which the first acts as the past transition path while the second serves as the evolution from current state to the next state, for example, Path 1 (State I→State IIa→State IIIa) and Path 2 (State I→State IIb→State I), which are schemed in Figure 4b. The State II is considered as the current state for both paths. Specifically, even at the same state, $K^+$ in mica have different memory effects due to their different accumulation pathways induced by the past states. Thus, for the second step, the same positive voltage can switch the device from State IIa to State IIIa or State IIb to State I. This feature suggests that the next state of the system not only relates to the current state, but also to the transport trail of the past states, which is defined as a typical non-Markov process. The experimental results in Figure 4c show the changing process of the resistance of the two examples. Table 1 summarizes all the possible non-Markov paths that can be achieved by the mica-based RRAM. Altogether, by using mica as the RRAM active layer, the non-Markov chain can be realized in a single device.

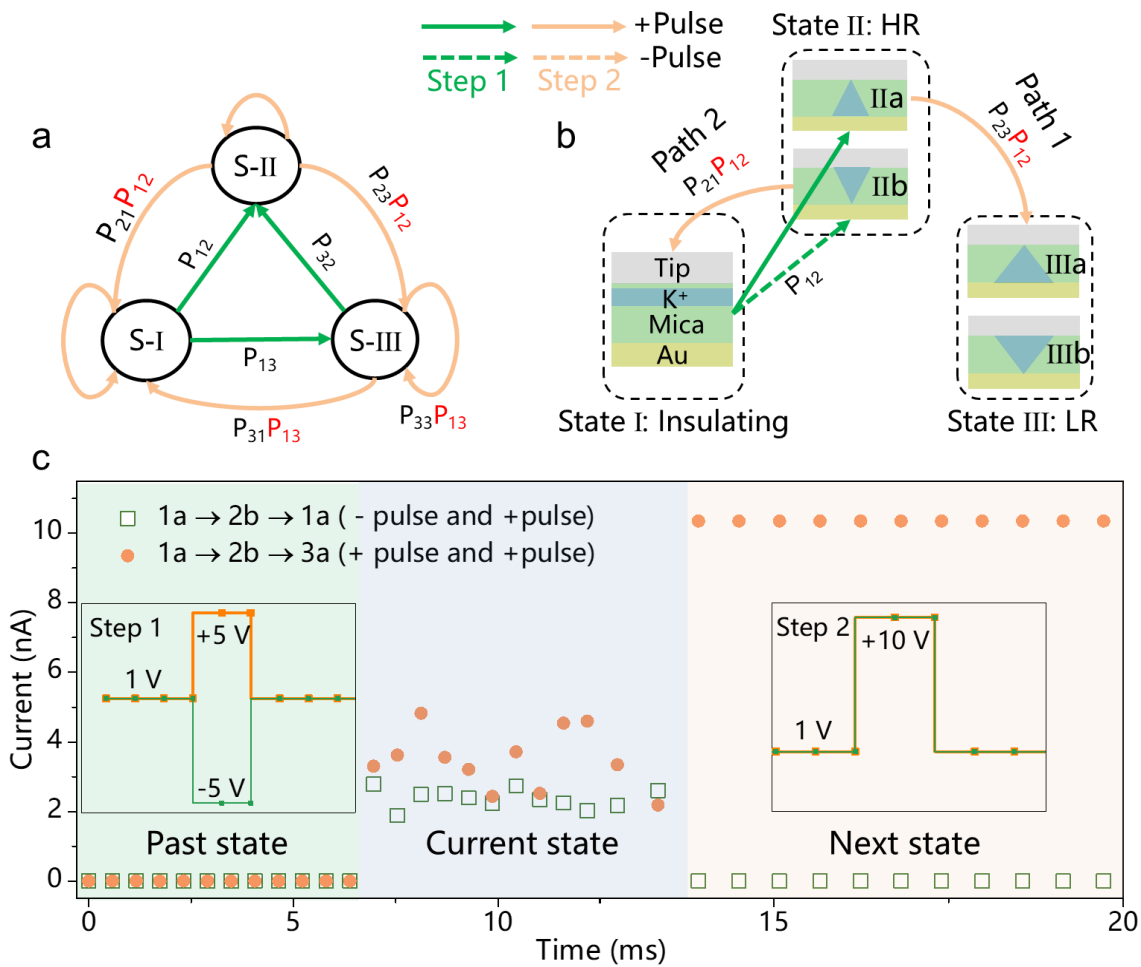

**Figure 4.** Realization of a non-Markov chain in a single RRAM. Schematics of (a) a typical non-Markov process and (b) a non-Markov chain achieved by a mica-based device. The three resistance states of the device are defined as State I, State II and State III, which correspond to insulating, HR and LR states, respectively. The applied positive and negative pulse voltages are indicated by solid and dotted arrows, respectively. The first and second step of the state transformation are represented by green and orange arrows, respectively. Examples of the non-Markov process by following the paths of (c) State I→State IIa→State IIIa and State I→State IIb→State I.

Table 1. Realization of all possible non-Markov paths in a single mica-based RRAM

| Comparison of two non-Markov paths | Input voltages |
|---|---|
| Path1: I→IIa→IIIa, path2: I→IIb→I | (5 V)→(10 V), (-5 V)→(10 V) |
| Path1: I→IIa→I, path2: I→IIb→IIIa | (5 V)→(-10 V), (-5 V)→(-10 V) |
| Path1: I→IIIa→IIIa, path2: I→IIIb→I | (10 V)→(10 V), (-10 V)→(10 V) |
| Path1: I→IIIa→I, path2: I→IIIb→IIIb | (10 V)→(-10 V), (-10 V)→(-10 V) |

**Conclusion**

We have achieved a non-Markov chain in a single 2D crystal RRAM device. The conductive mechanism of the device is related to the vertical migration of internal $K^+$ in the 2D mica under an electric field, which has been verified by both electrical transport characterization and depth profile with TOF-SIMS analysis. The accumulated $K^+$ show distinguished behaviors under electrical fields in opposite directions, resulting in the combination of unipolar and bipolar features of the 2D mica RRAM. Simultaneously, the device exhibits a high on/off ratio of $10^3$, good repeatability, and decent stability. The realization of a non-Markov chain in a single device opens a new horizon in highly-efficient and integrated systems toward future in-memory computing and intelligent devices. We envison that the scalable production of 2D mica and film assemblywould facilitate fabrication of large-scale device arrays for applications.

**Experimental Section**

*Device Fabrication*: For the mica-based RRAM, a Cr/Au (5 nm/50 nm) film was first deposited on a 300 nm-thick-SiO$_2$/Si substrate by an e-beam evaporation system (TSV-1500, Tianxingda Vacuum Coating Equipment Co., Ltd., China). Subsequently, mica flakes were mechanically exfoliated using scotch tape from the commercial bulk material (HQ graphene, Nederland) and directly transferred onto the Cr/Au-coated substrate.

Fabrication of probe/KI solution/mica/Au devices. First, PR (AZ 5214) was spin-coated (2000 rpm) to cover the whole substrate, followed by baking the sample on a hot plate at 125 °C for 1 min. Second, the shape of the bottom Au electrode was defined by a laser writer system (Dall, Aresis), followed by Au deposition using an e-beam evaporation system. Third, the mica was mechanically exfoliated by using scotch tape and transferred onto the bottom Au electrode with the assistance of PDMS using a homemade alignment station. Forth, another layer of PR was spin-coated and baked in the substrate. Fifth, laser writing process was performed to ablate a hole in the PR layer as a window to expose the mica flake to air. For device testing, a droplet of KI/DI water solution was dipped into the window.

Fabrication of graphite/mica/graphite and mica/MoS$_2$/graphite devices. First, the graphite (HQ graphene), muscovite mica (HQ graphene, Nederland), fluorophlogopite mica (Tiancheng Fluor phlogopite Mica Co., Ltd., China) and MoS$_2$ (HQ graphene, Nederland) flakes were exfoliated by using scotch tape from bulk crystal onto PDMS and identified by an optical microscope. Second, the target 2D flake was staked on a SiO$_2$ (300 nm)/Si substrate in a homemade alignment station with the assistance of PDMS. Third, PR (AZ 5214) was spin-coated (2000 rpm) to cover the whole substrate and baked on a hot plate (125 °C, 1 min). Forth, the electrodes were fabricated by the laser writer system and e-beam evaporation system.

*Electrical Measurements*: AFM characterization and CAFM measurements were performed by AFM (Cypher ES, Oxford Instruments, USA). For all the CAFM measurements, a fixed compliant current of 10 nA was applied. All the electrical measurements were done using a probe station, and relevant *I-V* curves were collected by a Keithley 4200 semiconductor parameter analyzer.

*The TOF-SIMS measurements*: The TOF-SIMS depth profiles were performed on a TOF-SIMS 5-100 instrument (IONTOF, Münster, Germany) equipped with a 30-keV $Bi_3^+$ liquid metal ion source for analysis and a 1-keV $Cs^+$ source for sputtering. Both sources striking the sample surface at an angle of 45°. $MCs^+$ mode measurements were carried out for detecting both $MCs^+$ ions (M is K, Al, Si) and $MCs_2^+$ (M is C, O) with analysis beam current of 0.7 pA and sputter beam current of 40 nA. Analysis area and sputter area were defined as 15×15 $\mu m^2$ and 200×200 $\mu m^2$, respectively. To achieve a better charge compensation during measurements, an electron flood gun and a non-interlaced scan mode were used.

**Acknowledgements**

R.Z. and W.C. contributed equally to this work. We thank the financial support from the National Natural Science Foundation of China (Nos. 51722206, 51920105002, 51991340, and 51991343), Guangdong Innovative and Entrepreneurial Research Team Program (No. 2017ZT07C341), and the Bureau of Industry and Information Technology of Shenzhen for the "2017 Graphene Manufacturing Innovation Center Project" (No. 201901171523).